\documentclass[aps,twocolumn,superscriptaddress,altaffilletter,lengthcheck, tightenlines,showpacs,showkeys]{revtex4}

\newcommand{\al}{\alpha}
\newcommand{\D}{\Delta}
\newcommand{\ben}{\begin{eqnarray}}
\newcommand{\een}{\end{eqnarray}}
\newcommand{\be}{\begin{equation}}
\newcommand{\ee}{\end{equation}}
\newcommand{\ba}{\begin{eqnarray}}
\newcommand{\ea}{\end{eqnarray}}
\newcommand{\n}{\label}
\newcommand{\no}{\noindent}
\newcommand{\la}{\lambda}
\newcommand{\La}{\Lambda}
\newcommand{\ga}{\gamma}
\newcommand{\ro}{\rho}

\newcommand{\Om}{\Omega}

\newcommand{\Dga}{\Delta\ga}
\newcommand{\bn}{\begin{equation}\label}

\usepackage{color}

\begin{document}

\title{Linear and nonlinear interactions in the dark sector\footnote{Dedicated to Professor Gilberto M. Kremer on the occasion of his sixtieth birthday.}}


\author{Luis P. Chimento}\email{chimento@df.uba.ar}
\affiliation{Department of Theoretical Physics, University of the Basque Country,
P.O. Box 644, 48080 Bilbao, Spain
and\\IKERBASQUE, the Basque Foundation for Science, 48011, Bilbao, Spain}

\affiliation{Departamento de F\'\i sica, Facultad
de Ciencias Exactas y Naturales, Universidad de Buenos Aires,
Ciudad Universitaria, Pabell\'on I, 1428 Buenos Aires,
Argentina,}

\bibliographystyle{plain}

\begin{abstract}
We investigate models of interacting dark matter and dark energy for the universe in a spatially flat Friedmann-Robertson-Walker (FRW) space-time. We find the "source equation" for the total energy density and determine the energy density of each dark component. We introduce an effective one-fluid description to evidence that interacting and unified models are related with each other, analyze the effective model and obtain the attractor solutions. We study linear and nonlinear interactions, the former comprises a linear combination of the dark matter and dark energy densities, their first derivatives, the total energy density, its first and second derivatives and a function of the scale factor. The latter is a possible generalization of the linear interaction consisting of an aggregate of the above linear combination and a significant nonlinear term built with a rational function of the dark matter and dark energy densities homogeneous of degree one. We solve the evolution equations of the dark components for both interactions and examine exhaustively several examples. There exist cases where the effective one-fluid description produces different alternatives to the $\La$CDM model and cases where the problem of coincidence is alleviated. In addition, we find that some nonlinear interactions yield an effective one-fluid model with a Chaplygin gas equation of state, whereas others generate  cosmological models with de Sitter and power-law expansions. We show that a generic nonlinear interaction induces an effective equation of state which depends on the scale factor in the same way that the variable modified Chaplygin gas model, giving rise to  the "relaxed Chaplygin gas model".

\end{abstract}
\vskip 1cm


\date{\today}
\maketitle

\section{Introduction}

Certain high precision astronomical observations suggest that the Universe entered an accelerated expansion stage when the value of its scale factor was approximately a half of the current one. This important discovery, which was based on the observations of the brightness of a class of supernovas (SNIa) \cite{Riess:1998cb}, has been confirmed by precise measurements of the spectrum of the cosmic microwave background (CMB) anisotropies \cite{Bennett} as well as the baryon acoustic oscillations (BAO) in the Sloan digital sky survey (SDSS) luminous galaxy sample \cite{Eisenstein:2005su}. This discovery has transformed Cosmology into a very active area in current Physics and will surely fix the bases of important advances in the future because the consensus between cosmologist points in the direction that the understanding of the phenomenon will probably require a unified comprehension of the gravitational and the other fundamental interactions. The above aspect of the expansion of the Universe, which becomes manifest on very large scales, will be only detectable from very distant astronomical objects.  

Basically, the evidences indicate that the Universe was dominated by nearly pressureless dark matter in the long lasting initial stage that goes back almost to its most early stages after the Big Bang. This was an epoch characterized by an every time slower expansion. However, the behavior got reversed and the Universe began an accelerate expansion under the domination of its dark energy component characterized by a negative pressure \cite{Sahni:1999gb}. This behavior has lasted till present and most likely  continue for ever. At large scales, there are strong evidences for a spatially flat and
accelerating universe  transiting from a scenario dominated by matter accumulated by purely attractive gravitational effects to another dominated by a dark component dispersed by repellent gravitational effects. The reasons, causes and details of when this transition happened, it was not still understood. These are part of the queries to answer in those projects which are framed in this active investigation area inside the Cosmology. It is interesting to note that for a sufficiently intense acceleration one can speak rather of superacceleration, then the possibility exists that the universe has a catastrophic end with a suddenly future singularity at a finite time (Big Rip) and a total disintegration of the well-known structures. 

To investigate the mechanisms that govern the dynamics of the evolution of the universe from its early stage until its recent accelerated phase we will consider fundamentally two types of model denominated respectively interacting and unified models. The aspects of the evolution in which we will be concerned are those connected to theoretical descriptions of the dark matter and dark energy.

In the interacting models, the source of Einstein equations which describe the dynamics of the universe  at large scale includes an aggregate of different material fluids and scalar fields that are conserved individually or interact among them. This is in principle the simplest, and perhaps the most obvious hypothesis, and it is in fact, the one that has provided more advances in the knowledge of the phenomenon of the recent acceleration of the universe \cite{Binder:2006yc}, \cite{waga}, and references therein.  Following observational evidences we will consider three fundamental components: baryons, dark matter and dark energy. Given the dynamical similarity between baryons and dark matter, we will make a simplified model replacing both components with a nearly pressureless dust, while the dark energy will be described by a fluid with a linear equation of state. This will allow us to focus our investigations on models of two fluids with energy transfer. Our goals will be the following ones: on one hand we will mainly investigate linear and nonlinear interactions, they will be considered as functions of  dark matter and dark energy densities, their first derivatives, the total energy density with its derivatives up to second order and the scale factor. On the other hand we will analyze the relation between interacting and unified models. We will also  examine the problem of coincidence: are the proportions of matter accumulated by gravitative effects and dark energy comparable at the present time for a strange coincidence or for a fundamental reason? Many models in the literature \cite{Matos} have been proposed to alleviate the problem of coincidence, for instance quintessence, k-essence, phantom, quintom, tachyon,   etc. 

In unified models, the Einstein equations will have a single component working as dark matter and dark energy at different stages. It interpolates smoothly between a matter dominated phase in the early stage and dark energy in the late stage of the evolution, so inducing an accelerated expansion of the universe. Consistently, the universe evolves from a power-law stage to a de Sitter stage. The Chaplygin gas and its extensions were the unified models that have been more studied in the literature \cite{Bilic}-\cite{lr}. Nevertheless, preliminary results indicate that observationally the equation of state of dark energy still can not be determined precisely. This information has produced several generalizations of those models, as for instance, the variable modified Chaplygin gas model with an equation of state, depending explicitly on the scale factor \cite{guo1}-\cite{deb}.

At the present, there exist some controversy between interacting and unified models that we can express in the following question, are unified models more probable and satisfactory than interacting models profusely studied in the literature? Perhaps, this controversy does not exist and one could expect some sort of resemblance between these different models. In this case it would be particularly interesting to find a relation between interacting and unified models. As far as we know, no study has been made in this direction. In the following we will examine, from the dynamical point of view, when interacting and unified models may be considered as similar ones.
  
The paper is organized as follows: in section II we consider two fluids, dark matter and dark energy, with energy transfer, develop an effective one-fluid description and find the source equation for the total energy density. Then, we write the evolution equation for the effective barotropic index, introduce a separable interaction and investigate the conditions of stability for constant solutions. In section III we introduce the "linear interaction I" and describe some interacting models particularly simple. After that we focus on the "linear interaction II" and find the exact scale factor and the effective barotropic index. Finally, we analyze a "general linear interaction" which induces a generalized $\La$CDM model. In section IV we consider a "nonlinear interaction" which includes a rational function of the energy densities of both dark components homogeneous of degree one and separate the analysis into two main cases. There, we show that the effective equation of state of two fluids with energy transfer includes the equations of state of the several generalizations given for the Chaplygin gas. In section V we give a prescription to obtain an interacting model from a unified one. Finally, in section VI the conclusions are stated.

\section{Dark sector evolution}

\subsection{Effective one-fluid description}

Let us consider an expanding universe modeled by a mixture of two interacting fluids, namely dark matter and dark energy with energy densities $\ro_c$ and $\ro_x$, and pressures $p_c$ and $p_x$ respectively. Due to the energy transfer between both dark components, they do not evolve separately \cite{Binder:2006yc} and the Einstein equations in a spatially flat FRW universe read
\ben
\label{00i}
&&3H^{2}=\ro_c+\ro_x,\\
\n{coi}
&&\dot\ro_c+\dot\ro_x+3H(\ro_c+p_c+\ro_x+p_x)=0.
\een

\no where $a$ is the scale factor and $H=\dot a/a$. The conservation equation (\ref{coi}) evidences the interaction among the components admitting the mutual exchange of energy and momentum. 

For the two dark components we assume  equations of state $p_c=(\ga_c-1)\ro_c$ and $p_x=(\ga_x-1)\ro_x$, where the barotropic indices $\ga_c$ and $\ga_x$ are constants. The dark matter is composed of nearly pressureless components with a barotropic index $\ga_c\approx 1$ and the dark energy has a barotropic index satisfying the condition $\ga_x<\ga_c$. Many of our results will even be valid when we include the possibility of phantom dark energy $\ga_x<0$. The total energy density $\ro$ and the conservation equation for the interacting two-fluid model are
\ben
\n{c1}
&&\ro=\ro_c+\ro_x,\\ 
\n{c2}
&&\ro'=-\ga_c\ro_c-\ga_x\ro_x,
\een
where the prime indicates differentiation with respect to the new time variable $'\equiv d/d\eta=d/3Hdt=d/d\ln{(a/a_0)^3}$ and $a_0$ is some value of reference for the scale factor. Solving the system of  equations (\ref{c1})-(\ref{c2}) we get the energy density of each dark component as a function of $\ro$ and its derivative $\ro'$ 
\be
\n{31}
\ro_c=-\frac{\ga_x\ro+\ro'}{\Dga\,}, \qquad \ro_x=\frac{\ga_c\ro+\ro'}{\Dga},
\ee
where $\Dga=\ga_c-\ga_x$ is the determinant of the linear equation system (\ref{c1})-(\ref{c2}), being positive for our model.  In turn, the ratio of the energies densities becomes $r=\ro_c/\ro_x=-(\ga_x\ro+\ro')/(\ga_c\ro+\ro')$. 

At this point, we introduce an energy transfer between the two fluids by separating the conservation equation for the system (\ref{c2}) into the two equations  
\ben
\n{de1}
&&\ro'_c+\ga_c\ro_c=-Q,\\
\n{de2}
&&\ro'_x+\ga_x\ro_x=Q.
\een
Here, we have consider a coupling with a factorized $H$ dependence $3HQ$ where the interaction term $Q$, with dimensions of an energy density, generates the energy transfer between the two fluids. With this assumption the dynamics of $\ro_c$ and $\ro_x$ is dictated by the scale factor instead of $H$.  Differentiating the first or the second Eq. (\ref{31}) and combining with the Eq. (\ref{de1}) or with the Eq. (\ref{de2}), we obtain a second order differential equation for the total energy density  
\be
\n{2} 
\rho''+(\ga_c+\ga_x)\rho'+\ga_c\ga_x\rho= Q\D\ga.
\ee
A similar equation was reported in \cite{Barrow} for the particular interaction $Q=c_1\ro_c+c_2\ro_x$, with $c_1$ and $c_2$ constants. 

The interacting two-fluid model has been reduced to an effective one-fluid model with total energy density $\ro$ and total pressure $p=p_c+p_x$, whose effective equation of state is
\be
\n{pe}
p(\ro,\ro')=-\ro-\ro'.
\ee 
From the above point of view and the conservation of the total energy-momentum tensor of the system, we assume an effective one-fluid description with equation of state $p=(\gamma-1)\rho$, where the effective barotropic index $\gamma=(\ga_c\ro_c+\ga_x\ro_x)/\rho$ ranges between $\ga_x<\ga<\ga_c$. The effective conservation equation becomes $\ro'+\ga\ro=0$, hence the expression for the energy density of the dark components (\ref{31}) are replaced by
\be
\n{ror}                   
\ro_c=-\frac{\ga_x-\ga}{\Dga}\,\ro,  \qquad  \ro_x=\frac{\ga_c-\ga}{\Dga}\,\ro,
\ee
and the ratio becomes $r=(\ga-\ga_x)/(\ga_c-\ga)$. Expressing these equations in terms of the energy density parameters 
\be
\n{OO}
\Omega_c= \frac{\ga-\ga_x}{\Dga},  \qquad \Omega_x=\frac{\ga_c-\ga}{\Dga},
\ee
we get $r=\Omega_c/\Omega_x$.

In other words, given an interaction $Q$, the total energy density $\ro$ of the effective one-fluid model is determined by  solving the source equation Eq. (\ref{2}). Once we know $\ro$, we are able to find the effective equation of state from Eq. (\ref{pe}) and the scale factor by integrating the Friedmann equation $3H^2=\rho$, without knowing $\ro_c$ and $\ro_x$ separately. Both energy densities are easily calculated by replacing $\ro$ and $\ro'$ into the Eq. (\ref{31}). For instance, in the no interaction case, $Q=0$, the  energy density of the effective one-fluid model is $\ro=c_{1}/a^{3\ga_c}+c_{2}/a^{3\ga_x}$, being $\ro_c$ the first term and $\ro_x$ the second one. Likewise $p=(\ga_c-1)\ro_c+(\ga_x-1)\ro_x=-\ro-\ro'$ is the effective equation of state. For any value of the constants $c_1$ and $c_2$ the total energy density $\ro\to c_{2}/a^{3\ga_x}$, the scale factor $a\to t^{2/3\ga_x}$ and the power-law solution  $t^{2/3\ga_x}$ becomes an attractor. Throughout the paper $c_i$, $c_i'$, .... and $b_i$, $b_i'$ .... with $i=0,1,2,3....$, will represent constants. 

Basically, we have shown that an interacting two-fluid model can be seen as an effective one-fluid model or equivalently considered as a unified one. Concerning this result, can the interacting two-fluid model or its unified version be derived from a Lagrangian? We observe that the dynamics of the unified or effective model is given by the two independent Einstein equations 
\ben
\label{00}
3H^{2}=\rho, \qquad
\dot\rho+3H(\rho+p)=0.
\een
These equations cannot determine the three quantities $a$, $p,$ and $\rho $ because we have one degree of freedom. Usually, the system of equations (\ref{00}) is closed with an equation of state $p=p(\rho)$. When we assume that the effective energy-momentum tensor $T_{ik}$ splits into two dark components, $T_{ik} =T_{ik}^c+ T_{ik}^x$, the Eqs. (\ref{00}) become Eqs. (\ref{00i}) and (\ref{coi}). The latter equations cannot determine the five quantities $a$, $\ro_c$, $\ro_x$, $p_c$ and $p_x$. To preserve the one degree of freedom of the unified model we have introduced an equation of state for each dark component $p_c=(\gamma_c-1)\rho_c$ and $p_x=(\gamma_x-1)\rho_x$. Then, by changing these equations of state we obtain a very large set of interacting models which are equivalent to a unified one, meaning that the decomposition into dark  matter and dark energy is not unique. We should take into account this degeneration and after that, we should go to the central point of deriving the interacting models from a Lagrangian. Some effort in this direction was reported in \cite{koivisto}.

Although in unified cosmologies we have not dealt with the evolution of large-scale inhomogeneities, it is important to mention something about its non-homogeneous generalizations. This is an issue of interest because candidates for the dark matter unification will only be valid if they ensure that initial perturbations can evolve into a deeply nonlinear regime to form a gravitational condensate of super-particles that can act like cold dark matter. In this sense one could follow the covariant and sufficiently general Zeldovich-like nonperturbative approach given in \cite{Bilic}. So that, 
it would be interesting to investigate in future works whether the equivalence between coupled and unified models holds also at the level of perturbation theory.

\subsection{Asymptotic stability}

The knowledge of stable solutions for the interacting two-fluid model is very useful because these attractor solutions determine the asymptotic behavior of the energy density of each dark component, so that, their contributions to the total energy density become constants. These scaling solutions are characterized by constant energy density parameters $\Omega_{cs}$, $\Omega_{xs}$ and they are reached for a broad range of initial conditions, hence alleviating the problem of coincidence. Other point of view consists in to look for a dynamical solution of the problem of coincidence such that the universe approaches to a stationary stage i.e., existence of attractor solutions for the evolution equation of the effective barotropic index. Thus, on the attractor, the ratio $r$ turns asymptotically constant, $r_s=\ro_{cs}/\ro_{xs}=\Omega_{cs}/\Omega_{xs}=(\ga_s-\ga_x)/(\ga_c-\ga_s)$. Consistently with both points of view the effective barotropic index tends to the asymptotic constant value $\ga_s$, as we can see from 
\be
\n{or}
\ga_s=\ga_c\Om_{cs}+\ga_x\Om_{xs}=\frac{r_s\ga_c+\ga_x}{1+r_s}.
\ee
From these two relations we can extract two different types of solutions ($i$) for $\ga_s\neq 0$, we integrate the barotropic index $\ga_s=-2\dot H/3H^2$ and obtain the power-law expansion $a=t^{2/3\ga_s}$, ($ii$) for $\ga_s=0$, we have a final de Sitter stage, $H= const$, with $\Om_{cs}$=$\Om_{xs}=0$ and $r_s=-\ga_x/\ga_c$. 

As Eq. (\ref{or}) relates the constant density parameters $\Om_{cs}$,  $\Om_{xs}$, the ratio $r_s$ and the effective barotropic index $\ga_s$, we will investigate the stability of the constant solution $\ga_s$ from the evolution equation of the effective barotropic index. Also, we will find the conditions of stability for the solutions $\ga_s$ and get the attractor when the energy transfer between both dark components is generated by a separable interaction. To this end, we deduce the differential equation for $\ga$ by differentiating $\ro'=-\ga\ro$ and by replacing $\ro'$ and $\ro''=(\ga^2-\ga')\ro$ into the source equation (\ref{2}) so, we have
\be
\n{41}
\ga'-(\ga-\ga_c)(\ga-\ga_x)=-\frac{\Dga}{\ro}\,\,Q.
\ee

Firstly, we assume that a constant solution $\ga=\ga_s$ of the Eq. (\ref{41}) with $\ga_x<\ga_s<\ga_c$  exists, and after that, we will impose the condition of stability so that $\ga_s$ be stable. An interaction satisfying the existence requirement belongs to the class 
\be
\n{qe}
Q(\ga_s)=\frac{(\ga_s-\ga_c)(\ga_s-\ga_x)}{\Dga}\,\ro,
\ee
with $Q(\ga_s)<0$. The negative value of $Q(\ga_s)$ indicates that the energy is being transferred from dark energy to dark matter, meaning that the latter component will dilute more slowly compared to its conserved evolution, $\ro_c\propto a^{-3\ga_c}$, whereas the accelerated expansion of the universe decreases compared with the noninteracting case, $\ga_x<\ga_s$. Interestingly enough, from Eq. (\ref{or}) the result $Q(\ga_s)<0$ guarantees that the ratio $r$ asymptotically tends to the constant value $r_s$, thus alleviating the problem of coincidence  \cite{gilberto}-\cite{Jackson}. For the class of interactions (\ref{qe}),  $\ga_s$ is a stationary solution of (\ref{41}) and we obtain the scale factor $a=t^{2/3\ga_s}$ by integrating $\ga_s=-2\dot H/3H^2$. 

In what follows the analysis of stability will be restricted to interactions that have the form $Q=Q(\ro_c,\ro_x,\ro_c',\ro_x',\ro,\ro',\ro'')$. By using the Eq. (\ref{31}) and $\ro'=-\ga\ro$ with $\ro''=(\ga^2-\ga')\ro$, we obtain that $\ro_{c,x}=\ro_{c,x}(\ro,\ro')$ and $\ro_{c,x}'=\ro_{c,x}'(\ro',\ro'')$, then the interaction becomes $Q=Q(\ga,\ga',\ro)$. For simplicity we  adopt separability of $Q$, that is, $Q=Q(\ga,\ga',\ro)=\ro \,Q(\ga,\ga')$ and write 
\be
\n{42}
Q(\ga,\ga',\ro)=\frac{(\ga-\ga_c)(\ga-\ga_x)}{\Dga}\,F(\ga,\ga')\ro ,
\ee
where the function $F$ depends on $\ga$ and $\ga'$. Several interacting models analyzed in the literature are described by the interaction term (\ref{42}), see for instance \cite{Barrow}-\cite{Valiviita}, \cite{Quercellini}. For later proposals the separable $Q$ also can be rewritten as
\be
\n{42'}
Q(\ga,\ga',\ro)=-\Dga\,\frac{\ro_c\ro_x}{\ro}\,F(\ga,\ga'),
\ee
by using the Eq. (\ref{ror}). In particular, for the $F=const.$ case, the latter interaction (\ref{42'}) will be investigated in detail in section IVB and there, we will show that the equation of state of the effective one-fluid model generates several versions of the modified Chaplygin gas equation of state. 

By combining Eqs. (\ref{41}) and (\ref{42}) we rewrite the evolution equation of the effective barotropic index as  
\be
\n{gaf}
\ga'=-(\ga-\ga_c)(\ga-\ga_x)(F-1).
\ee 
In conclusion, when the function $F$ fulfills the two conditions 
\be
\n{43}
F(\ga=\ga_s,\ga'=0)=1,
\ee
and 
\be
\n{cs}
\left(\frac{\partial\ga'}{\partial\ga}\right)_{(\ga_s,0)}=-\frac{(\ga_s-\ga_c)(\ga_s-\ga_x)F_{\ga}(\ga_s,0)}{1+(\ga_s-\ga_c)(\ga_s-\ga_x)F_{\ga'}(\ga_s,0)}<0,
\ee
where $F_{\ga}$ and $F_{\ga'}$ stand for the partial derivatives of $F$ with respect to $\ga$ and $\ga'$ respectively, then $\ga_s$ is a stable solution or an attractor. In other words, when the condition (\ref{43}) is satisfied the constant solution $\ga_s$ becomes a stationary solution of the  Eq. (\ref{gaf}). Besides, $\ga_s$ is stable whenever the condition of stability (\ref{cs}) is fulfilled. Hence the scaling solutions are attractors.

When the interaction $Q$ does not satisfies the existence requirement of belonging to the class (\ref{qe}) 
there is no constant solution of the evolution equation of $\ga$. However, in section IV we  will  examine the structural stability for some interactions which do not satisfy the existence requirement, as for instance, the inhomogeneous nonlinear interaction.

\section{Linear interaction}

Various cosmological models investigated in the literature are described by an interaction depending linearly on the energy densities $\ro_c$, $\ro_x$ and the total energy density $\ro$ \cite{Barrow}-\cite{Valiviita}. From the beginning we have seen in Eq. (\ref{31}), that $\ro_c$ and $\ro_x$ are linear functions of $\ro$ and its derivative $\ro'$. This encourage us to investigate linear combinations of $\ro_c$, $\ro_x$, $\ro$ and $\ro'$, namely the "linear interaction I" $(Q_l)$. We will study simple examples where the dark components interact with each other successively by only each term of $Q_l$ and review some of the models investigated with these particular couplings. For more generality,
we will extend this study to the case of considering the "linear interaction II" $(Q_L)$. It has new terms proportionals to the first derivative of the energy densities $\ro_c'$, $\ro_x'$ and a term proportional to the second derivative of the total energy density $\ro''$. It is motivated for the fact that the source equation (\ref{2}) becomes a linear second order differential equation for the interaction $Q_L$. In this extended case we will solve the Friedmann equation and find the exact scalar factor together with the effective barotropic index. Also, we consider the possibility of adding a constant term to $Q_L$ and  introduce the "general linear interaction" $(Q_{gL})$ with the intention of having a de Sitter scenario. After that we consider the contribution of a function of the scale factor and present the "more general linear interaction" $(Q_{mgL})$ to obtain other different final stages. Below, we will investigate the outstanding aspects of those interactions. 

\subsection{Linear interaction I}

Here we consider a linear combination of $\ro_c$, $\ro_x$, $\ro$ and $\ro'$ and define the "linear interaction I" in the following convenient form 
$$
Q_l(\ro_c,\ro_x,\ro,\ro')=c_1\frac{(\ga_s-\ga_c)(\ga_s-\ga_x)}{\Dga}\,\ro+c_2(\ga_s-\ga_c)\ro_c
$$
\be
\n{Ql}
-c_3(\ga_s-\ga_x)\ro_x-c_4\frac{(\ga_s-\ga_c)(\ga_s-\ga_x)}{\ga_s\Dga}\,\ro'.
\ee
It has been obtained from Eq. (\ref{42}) by choosing the function $F$,
$$F_l(\ga)=c_1\frac{(\ga_s-\ga_c)(\ga_s-\ga_x)}{(\ga-\ga_c)(\ga-\ga_x)}+c_2\frac{\ga_s-\ga_c}{\ga-\ga_c}$$
\be
\n{Fl}
+c_3\frac{\ga_s-\ga_x}{\ga-\ga_x}+c_4\frac{\ga(\ga_s-\ga_c)(\ga_s-\ga_x)}{\ga_s(\ga-\ga_c)(\ga-\ga_x)}.
\ee
Particular cases of the $Q_l$ (\ref{Ql}) with $c_1=c_4=0$ were previously investigated in Refs. \cite{Barrow}, \cite{Sadjadi}-\cite{Caldera}, with $c_2=c_3=c_4=0$ in \cite{zpc}-\cite{Wang}, with $c_1=c_3=c_4=0$ in \cite{Amendola} and with $c_1=c_2=c_4=0$ in \cite{gilberto}-\cite{Jackson}.

When we impose the condition (\ref{43}) to the function (\ref{Fl}), we obtain
\be
\n{con1}
c_1+c_2+c_3+c_4=1.
\ee
Then, a constant solution $\ga_s$ of the Eq. (\ref{41}) exists provided that the Eq. (\ref{con1}) be satisfied. Also, by combining Eqs. (\ref{31}), (\ref{Ql}) and (\ref{con1}), we reduce the interaction (\ref{Ql}) to a function which depends only on the total energy density and its first derivative
\ben
\n{Ql1}
&&Q_l(\ro,\ro')=\frac{h\ro+\ga_s^{-1}\left[h-(\ga_s-\ga_c)(\ga_s-\ga_x)\right]\ro'}{\Dga},\,\,\,\,\,\,\,\,\,\,\,\,\,\\
\nonumber
&&h=c_1(\ga_s-\ga_c)(\ga_s-\ga_x)-c_2\ga_x(\ga_s-\ga_c)\\
\n{s}
&&-c_3\ga_c(\ga_s-\ga_x).
\een 
Inserting this $Q_l$ into the Eq. (\ref{41}), we find the two constant solutions
\be
\n{2p}
\ga_l^-=\ga_s,   \qquad  \ga_l^+=\frac{\ga_c\ga_x-h}{\ga_s}.
\ee 
In turn, the condition of stability (\ref{cs}) gives
\be
\n{e-}
\ga_s-\ga_l^+<0,
\ee
so, the constant solution $\ga_s$ is asymptotically stable provided that  $\ga_s<\ga_l^+$ or $\ga_s^2<\ga_c\ga_x-h$. For positive energy densities also must be satisfied that $\ga_x<\ga_s<\ga_l^+<\ga_c$. Besides, for given values of $\ga_c$ and $\ga_x$, the inequality (\ref{e-}) bounds the range of constants $c_1$, $c_2$, $c_3$, $c_4$ and the form of the $Q_l$ (\ref{Ql}), yielding a stable cosmological model with the power-law expansion $a=t^{2/3\ga_s}$.

By writing the source equation (\ref{2}) for the $Q_l$ (\ref{Ql1}), we have that 
\be
\n{Ql''}
\ro_l''+\ga_s^{-1}\left(\ga_s^2+\ga_c\ga_x-h\right)\ro_l'+(\ga_c\ga_x-h)\ro_l=0,
\ee
and its general solution is 
\be
\n{sQl}
\ro_l=b_1a^{-3\ga_s}+b_2a^{-3\ga_l^+}.
\ee
This total energy density has a vanishing limit for an expanding universe. Then the interacting model is finally realized when the general solution (\ref{sQl}) is inserted into the energy density of each dark component (\ref{31}) and the effective equation of state (\ref{pe})
\ben
\n{r11}
&\ro_{cl}=\frac{\left(\ga_s-\ga_x\right)b_1a^{-3\ga_s}+\left(\ga_l^+-\ga_x\right)b_2a^{-3\ga_l^+}}{\Dga},\\
\n{r21}
&\ro_{xl}=\frac{\left(\ga_c-\ga_s\right)b_1a^{-3\ga_s}+\left(\ga_c-\ga_l^+\right)b_2a^{-3\ga_l^+}}{\Dga},\\
\n{p11}
&p_l=(\ga_s-1)\ro_l+(\ga_l^+-\ga_s)b_2a^{-3\ga_l^+}. 
\een
The coupling between the two dark components modifies dynamically, typical characteristics of $\ro_c$ and $\ro_x$. In fact, the universe begins with a mix of dark matter (\ref{r11}) and dark energy (\ref{r21}) represented approximately by the unstable energy densities $\ro_c \propto (\ga_l^+-\ga_x)a^{-3\ga_l^+}$ and $\ro_x\propto (\ga_c-\ga_l^+)a^{-3\ga_l^+}$ respectively. After that, the instability of the constant solution $\ga_l^+$ induces the universe to evolve from that unstable era, characterized by  $r_+=(\ga_l^+-\ga_x)/(\ga_c-\ga_l^+)$, to a stable final stage where the dark matter and dark energy densities are dominated by the stable components $\ro_c\propto (\ga_s-\ga_x)a^{-3\ga_s}$ and $\ro_x\propto (\ga_c-\ga_s)a^{-3\ga_s}$. The stable solution $\ga_s$ is associated to an asymptotically  stable ratio  $r_s=(\ga_s-\ga_x)/(\ga_c-\ga_s)$ with the expected result $r_+>r_s$, showing that the linear interaction alleviates the problem of coincidence. In turn, the scale factor interpolates between the unstable stage, evolving as  $a\propto t^{2/3\ga_l^+}$, and the stable stage evolving as $a\propto t^{2/3\ga_s}$. Meanwhile the effective equation of state  (\ref{p11}) plays the role of a peculiar fluid at the initial stage, while at late times it turns into the equation of state of a perfect fluid. For large scale factors the quantities $\ro$, $\ro_c$, $\ro_x$, $\ro'$, $p$ and $Q_l$ behave as  $a^{-3\ga_s}$, in this way, the original evolution equations (\ref{de1}) and (\ref{de2}) for the dark components become algebraic equations.

\subsection{Linear examples}

Now, we analyze four simple cases  by considering separately, each term of the $Q_l$ (\ref{Ql}). In all these examples we select the four function $F_l(\ga)$ in such a way that the condition (\ref{43}) is satisfied identically
\vskip .5cm
\no {\it 1.}\,\, $c_2=c_3=c_4=0$
\be
\n{c_1}
Q_\ro=\frac{(\ga_s-\ga_c)(\ga_s-\ga_x)}{\Dga}\,\ro,\quad F_\ro=\frac{(\ga_s-\ga_c)(\ga_s-\ga_x)}{(\ga-\ga_c)(\ga-\ga_x)}. \,\,\,\,\,\,\,\,\,
\ee
This negative interaction $Q_\ro<0$ represents an energy transfer from dark energy to dark matter. The interaction $Q_\ro$ between quintessence and a pressureless component in a spatially flat FRW cosmology produces a transition from a phase dominated by dark matter to an accelerated expansion phase dominated by dark energy \cite{zpc}-\cite{Wang}. Simultaneously the interaction $Q_\ro$ alleviates the problem of coincidence of the present universe \cite{Matos}. 

By imposing the condition (\ref{cs}) on the function $F_\ro$ (\ref{c_1}),
\be
\n{1'}
F_\ro'(\ga_s)=-\frac{1}{\ga_s-\ga_c}-\frac{1}{\ga_s-\ga_x}<0,
\ee
we find that the solution $\ga=\ga_s$ is an attractor provided $\ga_x<\ga_s<(\ga_c+\ga_x)/2<\ga_c$. Inserting $Q_\ro$ into the source equation (\ref{2}) we obtain the total energy density and the effective equation of state (\ref{pe}) 
\ben
\n{441}
&\ro=b_{1}a^{-3\ga_s}+b_{2}a^{-3(\ga_c+\ga_x-\ga_s)},\,\,\,\,\,\,\,\,\,\,\,\,\,\,\,\,\,\,\,\,\,\,\,\,\,\,\,\,\,\,\,\,\,\,\,\,\,\,\,\,\,\,\,\,\,\,\,\,\,\,\,\,\,\,\\
\n{pe1}
&p=(\ga_s-1)\ro+(\ga_c+\ga_x-2\ga_s)b_2a^{-3(\ga_c+\ga_x-\ga_s)}.\,\,\,\,\,\,\,\,\,
\een
For any value of the initial conditions $b_1$,$b_2$ and large scale factor, the total energy density $\ro\to b_{1}/a^{3\ga_s}$ and the effective equation of state (\ref{pe1}) adopts the barotropic perfect fluid form $p\approx (\ga_s-1)\ro$ with  $a\to t^{2/3\ga_s}$. Hence, the power-law expansion is asymptotically stable. 

The dark matter and dark energy densities (\ref{31}) can be written as
\ben
\n{re1}
&\ro_c=\frac{\left(\ga_s-\ga_x\right)b_1a^{-3\ga_s}+\left(\ga_c-\ga_s\right)b_2a^{-3(\ga_c+\ga_x-\ga_s)}}{\Dga},\\
\n{re21}
&\ro_x=\frac{\left(\ga_c-\ga_s\right)b_1a^{-3\ga_s}+\left(\ga_s-\ga_x\right)b_2a^{-3(\ga_c+\ga_x-\ga_s)}}{\Dga}.
\een
Thus, the ratio $r_{\ro}$ tends to $r_s= (\ga_s-\ga_x)/(\ga_c-\ga_s)$, being $r_s$  an attractor.

\vskip .5cm
\no \no {\it 2.}\,\, $c_1=c_3=c_4=0$
\be
\n{c_2}
Q_{\ro_c}=(\ga_s-\ga_c)\ro_c, \qquad F_{\ro_c}=\frac{\ga_s-\ga_c}{\ga-\ga_c}. 
\ee
For this interaction, with $Q_{\ro_c}<0$ \cite{Amendola}, the condition (\ref{cs}) is not satisfied 
\be
\n{2'}
F_{\ro_c}'(\ga_s)=(\ga_c-\ga_s)^{-1}>0,
\ee
and the power-law solution $a=t^{2/3\ga_s}$, with $\ga_s$ restricted to the interval $\ga_x<\ga_s<\ga_c$, is unstable. This model contains serious instabilities on the perturbations of the
dark energy component \cite{Valiviita}.

In this example the solution of the source equation (\ref{2}) with  the interaction $Q_{\ro_c}$ and the effective equation of state (\ref{pe}) are
\ben
\n{442}
&&\ro=b_{1}a^{-3\ga_s}+b_{2}a^{-3\ga_x},\\
\n{pe2}
&&p=(\ga_s-1)\ro+(\ga_x-\ga_s)b_2a^{-3\ga_x}.
\een
For any value of the initial conditions $b_1$,$b_2$ and for large scale factor, the total energy density (\ref{442}) has the limit $\ro_{\ro_c}\to b_{2}/a^{3\ga_x}$, meaning that $a\to t^{2/3\ga_x}$ since $\ga_x<\ga_s$. Accordingly, the effective equation of state (\ref{pe2}) becomes that of the dark energy $p\approx (\ga_x-1)\ro$ indicating that the interaction $Q_{\ro_c}$ would not be adequate to describe the evolution of dark components. The model seems to be completely dominated by the dark energy. In fact, the energy densities (\ref{31}) are
\ben
\n{re2}
&&\ro_c=\frac{\left(\ga_s-\ga_x\right)b_1a^{-3\ga_s}}{\Dga},\,\,\,\,\,\,\,\,\,\,\,\,\,\,\,\,\,\,\,\,\,\,\,\,\,\,\,\,\,\,\,\,\,\\
\n{re22}
&&\ro_x=\frac{\left(\ga_c-\ga_s\right)b_1a^{-3\ga_s}}{\Dga}+b_2a^{-3\ga_x},\,\,\,\,\,\,\,\,\,\,\,\,\,\,\,\,\,\,\,\,\,\,\,\,\,\,\,\,\,\,\,\,\,
\een
and the ratio $r_{\ro_c}\propto a^{-3(\ga_s-\ga_x)}\to 0$. Then, at late times, the interacting two-fluid model with energy transfer $Q_{\ro_c}$ would not solve the problem of coincidence and it would not be  suitable for to fit the present observations. However, this coupling can work when it is combined  linearly with some of the other parts of $Q_l$.

\vskip .5cm
\no \no {\it 3.}\,\, $c_1=c_2=c_4=0$
\be
\n{c3'}
Q_{\ro_x}=-(\ga_s-\ga_x)\ro_x, \qquad F_{\ro_x}=\frac{\ga_s-\ga_x}{\ga-\ga_x}, 
\ee
with $Q_{\ro_x}<0$. This interaction was examined in several papers \cite{gilberto}-\cite{Jackson}. Now the condition of stability (\ref{cs}) is satisfied 
\be
\n{3'}
F_{\ro_x}'(\ga_s)=(\ga_x-\ga_s)^{-1}<0,
\ee
and the solution $\ga_s$ is stable. By solving the source equation (\ref{2}) for $Q_{\ro_x}$ and using the Eq. (\ref{pe}), we obtain the total energy density and the effective equation of state, they read 
\ben
\n{443}
&&\ro=b_{1}a^{-3\ga_s}+b_{2}a^{-3\ga_c},\\
\n{pe3}
&&p=(\ga_s-1)\ro+(\ga_c-\ga_s)b_2a^{-3\ga_c}.
\een
For any value of the initial conditions $b_1$,$b_2$ and large scale factor, the total energy density $\ro\to c_{1}/a^{3\ga_s}$ and $a\to t^{2/3\ga_s}$ because $\ga_s<\ga_c$. The effective equation of state behaves as $p\approx (\ga_s-1)\ro$ showing that the interacting model is dominated by the attractor $\ga_s$.

The dark matter and dark energy densities (\ref{31}) are given by
\ben
\n{re3}
&&\ro_c=\frac{\left(\ga_s-\ga_x\right)b_1a^{-3\ga_s}}{\Dga}+b_2a^{-3\ga_c},\,\,\,\,\,\,\,\,\,\,\,\,\,\,\,\,\,\,\,\,\,\,\,\,\,\,\,\,\,\,\,\,\,\\
\n{re33}
&&\ro_x=\frac{\left(\ga_c-\ga_s\right)b_1a^{-3\ga_s}}{\Dga},\,\,\,\,\,\,\,\,\,\,\,\,\,\,\,\,\,\,\,\,\,\,\,\,\,\,\,\,\,\,\,\,\,
\een
showing that the ratio $r_{\ro_x}\to r_s=(\ga_s-\ga_x)/(\ga_c-\ga_s)$ on the attractor. Then, the interaction $Q_{\ro_x}$ may represent adequately a coupled model of dark matter and dark energy. In fact, the ratio of these components has enough parameters to be adapted to the observations, consequently this interacting model may be a candidate to alleviate the problem of coincidence. A cosmological model with the above characteristic was proposed for the current universe which consists of noninteracting baryonic matter and interacting dark components \cite{gilberto}. There it was used two interacting fluids in the dark sector with constant barotropic indices. The energy transfer was taken proportional to the dark energy density $Q_{\ro_x}$ and it was shown that the model leads to a correct behavior which is expected for a viable scenario of the present universe, as for instance, the deceleration parameter, density parameters and luminosity distance. Also the interaction $\ro_{\ro_x}$ was used to show that the overall energy transfer should go from dark energy to dark matter if the second law of thermodynamics and 
Le Ch\^{a}telier-Braun principle are to be fulfilled, guaranteeing that 
the ratio $r_{\ro_x}$ asymptotically tends to a constant, thus alleviating the problem of coincidence \cite{diego}. The evolution of a viscous cosmology model was also analyzed by employing an energy transfer between the dark components induced by the interaction $Q_{\ro_x}$ \cite{Chen}.

\vskip .5cm
\no \no {\it 4.}\,\, $c_1=c_2=c_3=0$
\be
\n{4'}
Q_{\ro'}=\frac{(\ga_c-\ga_s)(\ga_s-\ga_x)}{\ga_s\Dga}\,\ro', \quad
F_{\ro'}=\frac{\ga(\ga_s-\ga_c)(\ga_s-\ga_x)}{\ga_s(\ga-\ga_c)(\ga-\ga_x)}, 
\ee 
By using $\ro'=-\ga\ro<0$, we see that the interaction $Q_{\ro'}$ is negative. When the condition of stability (\ref{cs}) is imposed to the function $F_{\ro'}$, we get
\be
\n{4''}
F_{\ro'}'(\ga_s)=\frac{\ga_c\ga_x-\ga_s^2}{\ga_s(\ga_s-\ga_c)(\ga_s-\ga_x)}<0,
\ee
which means that $\ga_c\ga_x>\ga_s^2$. This inequality determines the range of values of the attractor $\ga_s$. Solving the source equation (\ref{2}) for $Q_{\ro'}$ and inserting it into the Eq. (\ref{pe}), we find the total energy density and the effective equation of state  
\ben
\n{444}
&&\ro=b_{1}a^{-3\ga_s}+b_{2}a^{-3\ga_c\ga_x/\ga_s},\\
\n{pe4}
&&p=(\ga_s-1)\ro+\ga_s^{-1}(\ga_c\ga_x-\ga_s^2)b_2a^{-3\ga_c\ga_x/\ga_s}. \,\,\,\,\,\,\,\,\,\,\,\,
\een
When $\ga_c\ga_x>\ga_s^2$ is satisfied whatever be the initial conditions $b_1$, $b_2$ the total energy density has the limit $\ro\to b_{1}/a^{3\ga_s}$ for large scale factor, evidencing that $\ga_s$ is an attractor.

The dark matter and dark energy densities (\ref{31}) are given by
\ben
\n{re4}
&\ro_c=\frac{\left(\ga_s-\ga_x\right)b_1a^{-3\ga_s}+\ga_x\ga_s^{-1}\left(\ga_c-\ga_s\right)b_2a^{-3\ga_c\ga_x/\ga_s}}{\Dga},\,\,\,\,\,\,\,\,\,\,\,\,\,\,\,\,\,\,\,\\
\n{re44}
&\ro_x=\frac{\left(\ga_c-\ga_s\right)b_1a^{-3\ga_s}+\ga_c\ga_s^{-1}\left(\ga_s-\ga_x\right)b_2a^{-3\ga_c\ga_x/\ga_s}}{\Dga},\,\,\,\,\,\,\,\,\,\,\,\,\,\,\,\,\,\,\,
\een
and $r_{\ro'}=(\ga_s-\ga_x)/(\ga_c-\ga_s)$ on the attractor. As far as we know, this interacting model was not investigated in the literature. It appears as a feasible candidate to be considered for to describe the evolution of the dark components and to alleviate the problem of coincidence.
    
\subsection{Linear interaction II}

To enlarge the set of linear interactions, we take into account a coupling with the first derivative of dark matter and dark energy densities, $\ro_c'$ and $\ro_x'$. Coming back to Eq. (\ref{31}), we observe that the above assumption introduces a dependence with the second derivative of the total energy density. Then, we generalize the $Q_l$ adding it these new terms and define the "linear interaction II", $Q_L$, in such a way that it verifies the conditions of stability (\ref{43})-(\ref{cs}). As far as we know, this kind of coupling was not investigated in the literature. So that, we will analyze in detail the $Q_L$.

We start building the $Q_L$ by combining linearly the quantities $\ro_c$, $\ro_x$, $\ro_c'$, $\ro_x'$, $\ro$, $\ro'$ and $\ro''$ 
\be
\n{lc}
Q_L=c_1\ro_c+c_2\ro_x+c_3\ro_c'+c_4\ro_x'+c_5\ro+c_6\ro'+c_7\ro''.
\ee 
By using the Eqs. (\ref{31}) the first and second terms of the $Q_L$ reduce to a linear combination of the total energy density and its derivative whereas the third and fourth terms reduce to a linear combination of the first and second derivatives of the total energy density. Then, rearranging all the terms in the Eq. (\ref{lc}), the $Q_L$ can be reduced to a linear combination of the basis elements $\ro$, $\ro'$ and $\ro''$. Finally we get
\be
\n{lcg}
Q_L=c_1'\ro+c_2'\ro'+c_3'\ro'',
\ee 
where the constants $c_i'$ are linear combinations of the constants $c_i$ in the $Q_L$ (\ref{lc}). On the other hand, the $Q_L$ (\ref{lcg}) also can be obtained from Eq. (\ref{42}) by selecting the  function $F_L(\ga,\ga')$ as 
\be
\n{fgl}
F_L=\frac{(\ga_s-\ga_c)(\ga_s-\ga_x)}{(\ga-\ga_c)(\ga-\ga_x)}\left[b_1+b_2\frac{\ga}{\ga_s}+b_3\frac{\ga^2-\ga'}{\ga_s^2}\right],
\ee
where we have used that $\ro'=-\ga\ro$, $\ro''=(\ga^2-\ga')\ro$. The coefficients
\ben
\n{bc'}
&&b_1=\frac{\Dga\, c_1'}{(\ga_s-\ga_c)(\ga_s-\ga_x)},\\
&&b_2=-\frac{\ga_s\Dga\, c_2'}{(\ga_s-\ga_c)(\ga_s-\ga_x)},\\
&&b_3=\frac{\ga_s^2\Dga\, c_3'}{(\ga_s-\ga_c)(\ga_s-\ga_x)},
\een
satisfy the constrain
\be\n{c4}
b_1+b_2+b_3=1,
\ee
after imposing the condition (\ref{43}) to the function (\ref{fgl}). By combining Eqs. (\ref{lcg}), (\ref{fgl}) and (\ref{c4}), we obtain the final $Q_L$ 
\be
\n{qlf}
Q_L=\frac{u\ro+\ga_s^{-1}\left[u+v-(\ga_s-\ga_c)(\ga_s-\ga_x)\right]\ro'+v\ga_s^{-2}\ro''}{\Dga},
\ee
with
\ben
u=(\ga_s-\ga_c)(\ga_s-\ga_x)b_1,\\
v=(\ga_s-\ga_c)(\ga_s-\ga_x)b_3.
\een
Coming back to the Eq. (\ref{41}), the $Q_L$ generates the two constant solutions
\be
\n{17}
\ga_L^-=\ga_s,    \qquad   \ga_L^+=\ga_s\frac{\ga_c\ga_x-u}{\ga_s^2-v},
\ee
while the condition of stability (\ref{cs}) yields
\be
\n{18}
\ga_s-\ga_L^+<0,
\ee
with the additional requirement $\ga_x<\ga_s<\ga_L^+<\ga_c$. 

For $Q_L$ (\ref{qlf}), the exact general solution of the source equation (\ref{2}) is given by
\be
\n{sQL}
\ro_L=b'_1a^{-3\ga_s}+b'_2a^{-3\ga_L^+}.
\ee
Then, the $Q_L$ induces a stable cosmological model with a final behavior described by the  power-law expansion, $a=t^{2/3\ga_s}$. In addition, the energy density of each dark component (\ref{31}) and the effective equation of state (\ref{pe}) are
\ben
\n{r12'}
&\ro_{cL}=\frac{\left(\ga_s-\ga_x\right)b'_1a^{-3\ga_s}+\left(\ga_L^+-\ga_x\right)b'_2a^{-3\ga_L^+}}{\Dga},\\
\n{r22}
&\ro_{xL}=\frac{\left(\ga_c-\ga_s\right)b'_1a^{-3\ga_s}+\left(\ga_c-\ga_L^+\right)b'_2a^{-3\ga_L^+}}{\Dga},\\
\n{p12}
&p_L=(\ga_s-1)\ro_L+(\ga_L^+-\ga_s)b'_2a^{-3\ga_L^+}.
\een
For large scale factors, similarly to the $Q_l$ case, the quantities $\ro$, $\ro_c$, $\ro_x$, $\ro_c'$, $\ro_x'$, $\ro'$, $\ro''$, $p$ and the $Q_L$ behave as  $a^{-3\ga_s}$, in this way, the original evolution equations of the dark components (\ref{de1}) and (\ref{de2}) become algebraic equations. In the initial regimen the above quantities behave as in the $Q_l$ case with $\ga_l^+$ substituted by $\ga_L^+$.

Finally, the Friedmann equation $3H^2=\ro_L$ for the source (\ref{sQL}) is implicitly solved and one finds the  scale factor 
\ben
\n{a}
&&a_L=\left[\omega\,\sinh{\Delta\tau}\right]^{2/3(\ga_L^+-\ga_s)},\,\,\,\,\,\,\,\,\,\,\,\,\,\,\,\,\,\,\,\,\,\,\,\,\,\,\,\,\,\,\,\,\,\,\,\,\,\\
\n{t}
&&t=\frac{2}{\sqrt{3b'_2}\,(\ga_L^+-\ga_s)}
\int \left[\omega\sinh{\Delta\tau}\right]^{\ga_s/(\ga_L^+-\ga_s)}d\tau,\,\,\,\,
\een
where $\omega^2=b'_1/b'_2$, see details in \cite{crossing}. Due to $\ga_L^+>\ga_s$, the latter equation shows that the variables $t$ and $\tau$ have the same asymptotic limits. Then, it is appropriate to investigate the scale factor and the remaining quantities in the two asymptotic regimes. 
The effective barotropic index reads
\be
\n{ge}
\ga_L=\frac{\ga_L^++\ga_s\sinh^2{\omega\Delta\tau}}{\cosh^2{\omega\Delta\tau}},
\ee
so as $t$ grows the model interpolates between the initial $\ga_L^+$ and the final $\ga_s$ values. Eqs. (\ref{a})-(\ref{ge}) allows us to express the dark matter and dark energy densities (\ref{ror}), the total energy density (\ref{sQL}), the ratio $r_L$ and the effective pressure (\ref{pe}) as functions of the new time $\tau$. In particular, at early and later times, the asymptotic limits of the ratio $r_L$ become
\be
\n{rl}
r_L^{+}=\frac{\ga_L^+-\ga_x}{\ga_c-\ga_L^+}, \qquad
r_{s}=\frac{\ga_s-\ga_x}{\ga_c-\ga_s}.
\ee
These ratios satisfy the crucial relation $r_L^{+}>r_{s}$, thus the $Q_L$ gives the possibility of alleviating the problem of coincidence. 

\subsection{General linear interaction and $\Lambda$CDM model }

We complete the subject of linear interaction by enlarging the basis elements with a constant, so that, the new base will be $c$, $\ro$, $\ro'$, $\ro''$. After that we generalize this basis elements by introducing a function of the scale factor instead of the constant $c$. Although, the effective one-fluid model is able to mimic the essential features of the $\Lambda$CDM cosmological model, clearly, the introduction of both  modifications could produce radical alternatives to the $\Lambda$CDM model.  

With this aim in mind, we first introduce the "general linear interaction", $Q_{gL}$, by adding the constant $Q_{0}/\Dga$ to the $Q_L$, so
\be
\n{gl}
Q_{gL}=\frac{Q_{0}}{\Dga}+Q_L,
\ee
where we have assumed that the constrain (\ref{c4}) holds for the constants in the $Q_L$. Obviously the condition of stability (\ref{18}) does not hold. A particular type of the linear combination (\ref{gl}), $Q=c_0+c_1\ro_c+c_2\ro_x$, was analyzed in \cite{Quercellini}.  Here, we investigate the consequences of the $Q_{gL}$ on the $\Lambda$CDM model. Combining Eqs. (\ref{2}), (\ref{qlf}), (\ref{17}) and (\ref{gl}) the source equation (\ref{2}) for the total energy density becomes
$$
(1-v\ga_s^{-2})\ro''+\ga_s^{-1}\left(\ga_s^2+\ga_c\ga_x-u-v\right)\ro'
$$
\be
\n{qgl}
+(\ga_c\ga_x-u)\ro=Q_{0}.
\ee
Its general solution and the corresponding effective equation of state (\ref{pe}), are
\ben
\n{sgl}
\ro_{gL}=\La_{eff}+b'_1a^{-3\ga_s}+b'_2a^{-3\ga_L^+},\,\,\,\,\,\,\,\,\,\,\,\,\,\,\,\,\,\,\,\,\,\,\,\,\,\,\,\,\,\,\,\,\,\,\,\,\,\,\,\,\,\,\,\,\,\,\,\,\,\,\,\\
\n{pgL}
p_{gL}=-\La_{eff}\ga_s+(\ga_s-1)\ro_{gL}+(\ga_L^+-\ga_s)b'_2a^{-3\ga_L^+},\,\,\,\,\,\,\,
\een
where $\La_{eff}=Q_{0}\ga_s/\ga_L^+(\ga_s^2-v)$ is the effective cosmological constant, induced by the constant term $Q_0/\Dga$ in the $Q_{gL}$. In order to $\La_{eff}>0$, we need to choose the parameters of the interacting model such that $Q_{0}/(\ga_s^2-v)>0$.

At late times the total energy density has the limit $\ro_{gL}\to\La_{eff}$ and the effective equation of state becomes $p\approx -\La_{eff}$. Thus, the effective one-fluid model can be associated with a unified dark sector model whose scale factor interpolates between a power-law phase and a de Sitter stage $H=H_0=\sqrt{3/\La_{eff}}$. Being $H_0$ the attractor solution in the phase space for any value of the initial conditions $b'_1$ and $b'_2$. 

From Eq. (\ref{31}), the dark matter and dark energy densities $\ro_{cL}$ and $\ro_{xL}$ are 
\be
\n{r1gl}
\ro_{cL}=-\frac{\ga_x\La_{eff}}{\Dga}+\ro_{cL}, \qquad \ro_{xL}=\frac{\ga_c\La_{eff}}{\Dga}+\ro_{xL}.
\ee
It is interesting to comment that, in general, when the total energy density tends asymptotically to $\La_{eff}$ for $a\to\infty$, the Eq. (\ref{31}) shows that the dark matter energy density has the final limit $\ro_c\to -\ga_x\La_{eff}/\Dga<0$. To relieve this problem we may assume a 
phantom equation state for the dark energy, with $\ga_x<0$, so the energy densities (\ref{r1gl}) become positive and the ratio $r_{gL}$ tends to $r_\infty=-\ga_x/\ga_c>0$. Curiously, the $Q_{gL}$ dresses the bare phantom dark energy density $\ro_x$ and gives a decreasing dark energy density (\ref{r1gl}) with the final limit $\ga_c\La_{eff}/\Dga$.  In the cases which we admit that $\ga_s$ or $\ga_L^+$, or both, are negatives one or the two dark components have a final phantom phase with increasing energy densities. 

A "more general linear interaction", $Q_{mgL}$, can be introduced by adding a well-behaved function $f(\eta)$ to the $Q_{gL}$, so we get 
\be
\n{mgl}
Q_{mgL}=\frac{\ga_L^+(\ga_s^2-v)}{\ga_s\Dga}\,\La_{eff}+\frac{f(\eta)}{\Dga}+Q_L.
\ee
The general solution of the source equation (\ref{2}) is obtained by finding the particular solution of 
$(1-v\ga_s^{-2})\ro''+\ga_s^{-1}\left(\ga_s^2+\ga_c\ga_x-u-v\right)\ro'+(\ga_c\ga_x-u)\ro=f(\eta)$ and adding it to the total energy density (\ref{sgl}). Essentially, for large scale factors, the behavior of the effective one-fluid model is defined by the relative weight between the constant $\ga_L^+(\ga_s^2-v)\La_{eff}/\ga_s\Dga$ and the function $f(\eta)$. In this way, the additional $f$ term determines the final behavior of the more general interaction model and how much it deviates from the cosmological $\Lambda$CDM model. The case where the coupling (\ref{mgl}) reduces to the $f$ term was investigated in \cite{Fabris}.

\section{nonlinear interaction}

Let us assume that the energy transfer between the dark matter and dark energy components is produced by the following "nonlinear interaction",
\be
\n{nl}
Q_{nL}=\frac{c_8\ro_c^2+c_9\ro_c\ro_x+c_{10}\ro_x^2}{\ro}+Q_L+\frac{f(\eta)\ro^\nu}{\Dga},
\ee
where the $Q_L$ is giving by Eq. (\ref{qlf}), $f(\eta)\ro^\nu$ is a nonlinear atypical term  proportional to a well-behaved function $f(\eta)$, which depends on the scale factor $\eta=\ln{a^3}$, and to $\ro^\nu$. The constant $\nu$ will be determined later on so that the source equation (\ref{2}) can be recast in a solvable equation. 

So far we have examined interactions depending linearly on: the dark matter and dark energy densities, their first derivatives, the total energy density and its derivatives up to second order. However, it will be interesting to consider an interaction term which includes a rational function of the energy densities $\ro_c$ and $\ro_x$ homogeneous of degree one, as for instance the first three terms in the RHS of the Eq. (\ref{nl}). It could be considered as a possible generalization of the $Q_l$ (\ref{Ql1}) and the $Q_L$ (\ref{qlf}). In subsection IVB we will see that the rational interaction terms are of particular importance. Besides, a coupling depending on the scale factor will be useful so that the equations of state of the effective one-fluid model produces equations of state which extend and generalize those of the Chaplygin gas \cite{Bilic}-\cite{lr} and the variable modified Chaplygin gas model \cite{guo1}-\cite{deb}. 

By using Eq. (\ref{31}) we have shown that the $Q_L$ reduces to a linear combination of the three elements of the base $\ro=c_1'\ro+c_2'\ro'+c_3'\ro''$, (see Eqs. (\ref{lcg}) and (\ref{qlf})). Following similar arguments, the three terms in the numerator of the nonlinear part of the Eq. (\ref{nl}), $c_8\ro_c^2+c_9\ro_c\ro_x+c_{10}\ro_x^2$, become a linear combination of $\ro^2$, $\ro\ro'$ and $\ro'^2$. Rearranging all these terms, we obtain the final form of the nonlinear interaction (\ref{nl}),
\be
\n{final}
Q_{nL}=\frac{c_1''\ro'^2+\ro(c_2''\ro+c_3''\ro'+c_4''\ro'')+f(\eta)\ro^{\nu+1}}{\ro\Dga},
\ee
where the $c_i''$ are linear combinations of the remaining constants. Below we will see that the nonlinear terms arising from the homogeneous function of degree one in the energy densities $\ro_c$, $\ro_x$ and $f(\eta)\ro^\nu$ produce original results, as for instance, the effective equation of state describes universes which have power-law and the Sitter expansions. Also these terms are really the responsible that the effective equation of state has a reminiscence of the Chaplygin gas, including its different versions.  
From Eqs. (\ref{2}) and (\ref{final}), we obtain the nonlinear differential equation for $\ro$
$$
\ro\ro''+\frac{\ga_c+\ga_x-c_3''}{1-c_4''}\ro\ro'-\frac{c_1''}{1-c_4''}\ro'^2\,\,\,\,\,\,\,\,\,\,\,\,\,\,\,\,\,\,\,\,\,\,\,\,\,\,\,\,\,\,\,
$$
\be
\n{Qnl}
+\frac{\ga_c\ga_x-c_2''}{1-c_4''}\ro^2=\frac{f}{1-c_4''}\ro^{\nu+1}.
\ee
In the particular case where the first and last terms in the RHS of the $Q_{nL}$ (\ref{final}) vanish simultaneously, the Eq. (\ref{Qnl}) becomes a homogeneous linear differential equation for $\ro$. However, in other cases, the general solution of the source equation (\ref{Qnl}) will be obtained from a nonlinear superposition of the two basis solutions of a second order linear differential equation. 

Renaming the four constants in the Eq. (\ref{Qnl}) by $b_1$, $b_2$, $b_3$ and $b_4$ respectively (from left to right) and changing to the new variable $x=\ro^{(1+b_2)}$ with $b_2\neq -1$ or $c_1''+c_4''\neq 1$, the Eq. (\ref{Qnl}) turns into the equation of a forced dissipative $(b_1>0)$ or antidissipative $(b_1<0)$ linear oscillator
\be
\n{x''}
x''+b_1x'+b_3(1+b_2)x=b_4(1+b_2)f(\eta), 
\ee
where we have chosen $\nu=-b_2=c_1''/(1-c_4'')$. If $x_{1h}$ and $x_{2h}$ are the two basis solutions of the homogeneous Eq. (\ref{x''}), then the general solutions of the Eq. (\ref{Qnl}) can be written as a nonlinear superposition of these basis solutions $\ro_{nL}=(b_5x_{1h}+b_6x_{2h}+x_p)^{1/(1+b_2)}$. There $x_p$ is the particular solution of the Eq. (\ref{x''}) and $b_5$, $b_6$ are integration constants. From now on, the Eq. (\ref{x''}) will substitute the source equation (\ref{2})  for the $Q_{nL}$.

In what follows we will study the solution of the Eq. (\ref{x''}) considering separately: the homogeneous case $f=0$, the example of Chaplygin gas and the inhomogeneous case $f\neq 0$. The latter it will be divided into two parts, the $f=f_0=const.$ case, and the $f=f(a)$ case where the "relaxed Chaplygin gas model" will emerge. 

\subsection{The homogeneous case}

In the $f=0$ case, the source equation Eq. (\ref{x''}) becomes homogeneous and the $Q_{nL}$ (\ref{final}) reduces to
\be
\n{nlh}
Q_h=\frac{c_1'\ro^{-1}\ro'^2+c_2'\ro+c_3'\ro'+c_4'\ro''}{\Dga}.
\ee
By inserting the "homogeneous nonlinear interaction" ($Q_h$) into the Eq. (\ref{x''}), we easily get the general solution $x_h=b_5a^{3\la^-_{nL}}+b_6a^{3\la^+_{nL}}$. Then, coming back to original variable $\ro_h=x_h^{1/(1+b_2)}$, we have the total energy density, it reads
\be
\n{sg}
\ro_h=\left[b_5a^{3\la^-_{nL}}+b_6a^{3\la^+_{nL}}\right]^{1/(1+b_2)},
\ee
where $b_5$ and $b_6$ are integration constants while $\la^-_{nL}$ and $\la^+_{nL}$ are the characteristic roots of the linear source equation (\ref{x''})
\be
\n{char}
\la_{nL}^\mp=\frac{-b_1\mp\sqrt{b_1^2-4b_3(1+b_2)}}{2}.
\ee
On the other hand, the dark matter and dark energy densities (\ref{31}) are
\ben
\n{r1nl}
\ro_{ch}=-D\left[[\la^-_{nL}+\ga_x(1+b_2)]\ro+b_6\D\la\,\,\frac{a^{3\la^+_{nL}}}{\ro^{b_2}}\right],\,\,\,\,\,\\
\ro_{xh}=D\left[[\la^-_{nL}+\ga_c(1+b_2)]\ro+b_6\D\la\,\,\frac{a^{3\la^+_{nL}}}{\ro^{b_2}}\right],\,\,\,\,\,\,\,\,\,\,\,
\een
where $D=[(1+b_2)\Dga]^{-1}$ and $\D\la=\la^+_{nL}-\la^-_{nL}$.

We have started from an interacting two-fluid model and finally obtained an unified cosmological model where the dark matter and the dark energy evolve as an effective one-fluid. Now, combining Eqs. (\ref{pe}) and (\ref{sg}), we find the effective equation of state  
\be
\n{e1}
p_h=-\left(1+\frac{\la^-_{nL}}{1+b_2}\right)\ro_h-\frac{b_6\D\la\,\,a^{3\la^+_{nL}}}{(1+b_2)\ro_h^{b_2}}.
\ee

Depending on the values of the parameters $b_1$, $b_2$, and $b_3$ the total energy density (\ref{sg}) behaves asymptotically as $\ro_h\to a^{3\la^\pm_{nL}/(1+b_2)}$ in the limit of large scale factors, meaning that $a\to t^{-2(1+b_2)/3\la^\pm_{nL}}$. For $b_5=0$ (+) or $b_6=0$ (-), the model includes the exact power-law expansions $a^\pm=t^{-2(1+b_2)/3\la^\pm_{nL}}$. In particular, when $(1+b_2)/3\la^\pm_{nL}>0$, we have a final phantom phase entailing that the scale factor expands so quickly that the scalar curvature $R\to\infty$ in the limit $a\to\infty$ and it reaches $a=\infty$ in a finite amount of proper time. In the special cases that $\la^\pm_{nL}=0$ and $\la^\mp_{nL}<0$, namely $b_1>0$ and $b_3=0$ or $b_1<0$ and $b_3=0$ (see Eq. (\ref{char})), we have a final de Sitter stage with an effective cosmological constant given by the limit $\ro_h\to\La_{eff}=b_6^{1/(1+b_2)}$ or $\ro_h\to\La_{eff}=b_5^{1/(1+b_2)}$. These models include the modified Chaplygin gas introduced in Ref. \cite{marian} where it was proposed the equation of state $p=A\ro-B/\ro^n$, with $n\geq 1$, and the parameters $A$ and $B$ were constrained to be positive. When both, $\la^+_{nL}\neq 0$ and $\la^-_{nL}\neq 0$, the equation of state (\ref{e1}) contains those which characterize various of the variable modified Chaplygin gas models investigated in Refs.  \cite{guo1}-\cite{deb}. In the next subsection we concentrate on a specific nonlinear interaction and study the two associated cases where the characteristic root $\la^+_{nL}$ vanishes. 

\subsection{Nonlinear examples}

Here we investigate the interesting case where $Q_{h}$ takes the form
\be
\n{ch}
Q_{h0}=\al\ga_c\frac{\ro_c\ro_x}{\ro},
\ee
with $\al$ constant, see \cite{waga} and references therein. The interaction (\ref{ch}) also can be obtained from the Eq. (\ref{42'}) by choosing the constant function  $F(\ga)=-\al \ga_c/\Dga$. From Eqs. (\ref{Qnl}) and (\ref{ch})  we identify the three coefficients $b_1=(\ga_c+\ga_x)(1+b_2)$, $b_2=\al\ga_c/\Dga$ and $b_3=\ga_c\ga_x(1+b_2)$. Then, the characteristic roots (\ref{char}) of the source equation (\ref{x''}) read
\bn{r+-}
\la_{nL}^+=-\ga_x(1+b_2),  \qquad  \la_{nL}^-=-\ga_c(1+b_2).
\ee
Introducing these roots into the homogeneous general solution (\ref{sg}) and the effective equation of state (\ref{e1}), they reduce to 
\ben
\n{scha}
&&\ro=\left[b_5a^{-3\ga_c(1+b_2)}+b_6a^{-3\ga_x(1+b_2)}\right]^{1/(1+b_2)},\\
&&p=(\ga_c-1)\ro-b_6\Dga\frac{ a^{-3\ga_x(1+b_2)}}{\ro^{b_2}}.
\een
This effective equation of state can be identified with those which were used to build variable modified Chaplygin gas models \cite{guo1}-\cite{deb}.

\vskip .5cm
\no {\it 1. modified Chaplygin gas}
\vskip .5cm

We make an adequate selection of the parameters so that the interaction (\ref{ch}) be focused on a dark energy component described by some kind of vacuum energy density i.e., $\ga_x=0$. Then, $b_3=\ga_c\ga_x(1+b_2)$=0, $\D\ga=\ga_c$, $b_2=\al\ne-1$, $\la^+_{nL}=0$, and the second term in Eq. (\ref{scha}) becomes constant. Assuming that the dark matter component is nearly pressureless, we may associate it to a barotropic fluid with a free constant parameter $\ga_c=\ga_0\approx 1$. Then the energy density (\ref{scha}) abbreviates to
\be
\n{rocha}
\rho=\left[\frac{B}{\ga_0}\pm\left(\frac{a_0}{a}\right)^{3{\ga_0(1+\alpha)}}\right]^{{1}/{1+\alpha}},
\label{r}\ee
where the new constants $B$ and $a_0$ are redefinitions of the old integration constants $b_5$ and $b_6$. Hence, by replacing this energy density in Eq. (\ref{pe}), we obtain the equation of state of the one effective fluid
\be
\n{pcha}
p=(\ga_0-1)\rho-\frac{B}{\rho^{\alpha}}\label{p}.
\ee
It characterizes several unified cosmologies implemented with Chaplygin gases  as the generalized, extended, modified and enlarged ones \cite{Bilic}-\cite{lr}. Also, these unified models along with some others ones \cite{arm}-\cite{log2}, were extensively used to describe unified versions of dark matter and dark energy. 

\vskip .5cm
\no {\it 2. reduced unified model}
\vskip .5cm

For $\ga_0=0$, the effective one-fluid model described by the expressions (\ref{rocha}) and (\ref{pcha}) is not valid. In this "reduced unified model" the source equation (\ref{x''}) must be solved again by making the particular choices $b_1=b_3=0$ and $b_2=\al$. After comparing the latter particular choices  with the coefficients of the  nonlinear version of the source equation (\ref{Qnl}) we obtain the coefficients of the corresponding "reduced nonlinear interaction" $(Q_r)$, $c_3''=\ga_c+\ga_x$, $c_2''=\ga_c\ga_x$, $f=0$ and $c_1''=\al(c_4''-1)$ respectively. Then, from Eqs. (\ref{nl}) and (\ref{final}), we get the final $Q_r$  
\bn{Qpoly}
Q_{r}=\frac{\al(c_4''-1)\ro^{-1}\ro'^2+\ga_c\ga_x\ro+(\ga_c+\ga_x)\ro'+c_4''\ro''}{\Dga}.
\ee
Now, the source equation (\ref{x''}) reduces to $x''=0$ and its general solution is $x=b_8+b_9\eta$. Thus, one finds that the total energy density $\ro=x^{1/(1+\al)}$ has a logarithmic dependence with the scale factor
\bn{poly}
\ro=\ro_0\left[\pm 1+b\ln{\left(\frac{a}{a_0}\right)^3}\right]^{1/(1+\al)},
\ee
where the constants $\ro_0$ and $b$ are related to the old integration constants $b_8$ and $b_9$. In turn by using Eqs.  (\ref{31}) and (\ref{pe}), we find the dark matter and dark energy densities, and the equation of state of the effective one-fluid model are
\ben
\n{polyp}
&&\ro_{c}=-\frac{\ro}{\Dga}\left[\ga_x+\frac{b}{1+\al}\,\left(\frac{\ro_0}{\ro}\right)^{1+\al}\,\right],\\
&&\ro_{x}=\frac{\ro}{\Dga}\left[\ga_c+\frac{b}{1+\al}\,\left(\frac{\ro_0}{\ro}\right)^{1+\al}\,\right],\\
&&p=-\ro-\frac{b\ro_0}{1+\al}\,\,\left(\frac{\ro_0}{\ro}\right)^{\al}.
\een
The last equation of state also can be seen as a generalization of the polytropic equation of state $p=K\ro^{\ga_p}$, where $K$ is a constant and $\ga_p$ is the polytropic index. On the other hand, by integrating the Friedmann equation with the total energy density (\ref{poly}), we determine the exact scale factor 
\bn{spoly}
a=a_0\exp{\frac{1}{3b}\left[\pm 1+\left[\frac{b\sqrt{3\ro_0}(2\al+1)\,t}{2(1+\al)}\right]^{\frac{2(1+\al)}{(2\al+1)}}\right]},\,\,\,\,\,\,\,\,\,\,\,\,\,
\ee
where we have set $t_0=0$. Note that the final results (\ref{poly})-(\ref{spoly}) are independent of $c_4''$, then the interaction term $\ro''$ dos not contribute to the evolution of this reduced unified model. 

\subsection{The inhomogeneous case and the "relaxed Chaplygin gas model"}

Here we are going to consider the "inhomogeneous nonlinear interaction" $(Q_i)$,   
\be
\n{final'}
Q_{i}=\frac{c_1''\ro^{-1}\ro'^2+c_2''\ro+c_3''\ro'+c_4''\ro''+f(\eta)\ro^{c_1''/(1-c_4'')}}{\Dga},
\ee
which has been obtained from the Eq. (\ref{final}) by choosing $\nu=-b_2=c_1''/(1-c_4'')$. The source equation (\ref{x''}) becomes inhomogeneous and the general solution is obtained by adding a particular solution $x_p$ to the homogeneous solution $x_h=b_5a^{3\la^-_{nL}}+b_6a^{3\la^+_{nL}}$, then the general solution $\ro_{i}=(x_h+x_p)^{1/(1+b_2)}$ of the Eq. \ref{x''}) reads
\be
\n{sgr}
\ro_{i}=\left[b_5a^{3\la^-_{nL}}+b_6a^{3\la^+_{nL}}+x_p\right]^{1/(1+b_2)},
\ee
and the equation of state (\ref{pe}) takes the form
\be
\n{pgr}
p_{i}=-\left[1+\frac{\la^-_{nL}}{1+b_2}\right]\ro_{i}-\frac{b_6\D\la\,\,a^{3\la^+_{nL}}+x_p'-\la^-_{nL}x_p}{(1+b_2)\ro_{i}^{b_2}}.
\ee

${(i)}$ For $f=f_0=const.$ the particular solution $x_p=b_4f_0/b_3$ is constant and the total energy density (\ref{sgr}) along with the effective equation of state (\ref{pgr}) describe a "double unified model", in a sense that initially the universe is dominated by the "two" terms inside the bracket of $\ro_{i}\approx (b_5a^{3\la^-_{nL}}+b_6a^{3\la^+_{nL}})^{1/(1+b_2)}$. These "two" terms can be seen as a "nonlinear mixture of two fluids". But at late times the universe is dominated by a vacuum energy $\ro_{i}\approx x_p^{1/(1+b_2)}$ and has a de Sitter expansion. So, this ${(i)}$ case includes various generalizations of the modified Chaplygin gas model investigated in Refs. \cite{guo1}-\cite{deb}. When the constants $b_5$ or $b_6$ vanishes the effective equation of state (\ref{pgr}) turns in Eq. (\ref{pcha}) and the double unified model produces different versions of the Chaplygin gas \cite{marian}-\cite{lr}.

${(ii)}$ For $f(\eta)\ne const.$ and $\eta=\ln{a}$, the particular solution $x_p$ and consequently the numerator of the last term in the RHS of the effective equation of state (\ref{pgr}) become arbitrary functions of the scale factor. Then, a fluid obeying the Eq. (\ref{pgr}) defines a "relaxed Chaplygin gas model".

Finally, we investigate the structural stability of the solutions corresponding to the source equation (\ref{x''}) for a well-behaved function $f(\eta)$, which will be considered as a degree of freedom in the problem of stability. 

We use the analogy with the classical potential problem by writing the Eq. (\ref{x''}) as an equation of motion for a dissipative mechanical system, namely,
\be
\n{em}
\frac{d}{d\eta}\left[\frac{x'^2}{2}+V(x,\eta)\right]=-D(x,x',\eta),
\ee
where
\ben
\n{V}
&&V(x,\eta)=\frac{b_3(1+b_2)}{2}\left[\left(x-\frac{b_4}{b_3}f\right)^2-\frac{b_4^2}{b_3^2}f^2\right],\,\,\,\,\,\,\,\,\,\,\,\,\,\\
\n{D}
&&D(x,x',\eta)=b_4(1+b_2)f'x+b_1x'^2,
\een
$V(x,\eta)$ is the "potential" and $D(x,x',\eta)$ is the "dissipation" of the "equivalent mechanical system". In addition, we assume that the potential $V$ is inferiorly bounded by a function $V_b$, it has a finite limit for $\eta\to\infty$ and the dissipation $D$ is positive definite in the same limit. 

When $b_3(1+b_2)>0$, the potential becomes bounded by the function $V_b$, thus $V\ge V_b$, where 
\bn{Vm}
V_b=-\frac{b_4^2(1+b_2)}{2b_3}\,f^2.
\ee
This bound is obtained by evaluating the potential (\ref{V}) on the limit solution $x_{lim}=b_4f/b_3$ of the source equation (\ref{x''}) for $\eta\to\infty$, or $\ro_{lim}=(b_4f/b_3)^{1/(1+b_2)}$. Therefore, for a dissipative mechanical system with $b_1>0$, the set of functions $f(\eta)$ such that $f\to f_0$ for $\eta\to\infty$ and $b_4(1+b_2)f'>0$ define a positive definite dissipation, $D>0$. In this context,  a solution of the source equation which have the limit $\ro\to\ro_{min}$ may be considered as a "stable solution".

\section{Starting with unified models and ending with interacting ones}

Eqs.~(\ref{31})-(\ref{de2}) introduce an alternative interpretation of unified models by associating them with interacting models. Several kinds of unified models were extensively investigated in the literature \cite{arm}-\cite{log2}, for instance: the Chaplygin gas and purely kinetic, quintessence, k-essence and DBI cosmologies. For purely kinetic cosmologies we mean that the pressure and energy density of each field involve only the time derivative of the same field. 

At this point we consider two options, in the first one the equation of state of the unified model has the form $p=p(\ro)$, as it is for the Chaplygin gas (see the Eq. (\ref{pcha})) or we have  purely kinetic quintessence, k-essence and DBI fields. These are described by the equations of state $p_q=\ro_q-2V_0$, $p_k=-{\cal F}(\dot\phi^2(\ro_k))$ where ${\cal F}={\cal F}(\dot\phi^2)$ is the kinetic function and $\phi$ is the k-essence field, and $p_{DBI}=-V_0+(\rho-V_0)/(\rho_{DBI} f_0-V_0f_0+1)$ for Dirac-Born-Infeld (DBI) cosmologies \cite{log2}. The parameter $V_0$ represents a constant potential for the three fields and $f_0$ is the constant warp factor of the metric. In all the above unified cosmological models, after integrating the conservation equation of each component, the total energy density depends only on the scale factor, $\ro=\ro(a)$. In the second option we include the cases where the equation of state $p=p(\ro)$ is given but we do not know $\ro=\ro(a)$ explicitly.

To clarify the alternative interpretation, let us assume that we have found the energy density $\ro=\ro(a)$ of the fluid which characterizes a unified model. Then the Eq. (\ref{31}) associates this unified model with a two-fluid model whose components have energy densities $\ro_c(a)$ and $\ro_x(a)$ respectively. Besides, inserting $\ro(a)$ into the source equation (\ref{2}), we can calculate the interaction $Q=Q(a)$. Therefore, a one-fluid model with energy density $\ro(a)$ interpolating between dark matter and dark energy scenarios, can be expressed as a two-fluid model with energy transfer. In other words they share the same Friedmann equation and the same scale factor, so they define the same geometry.

In the case we know the equation of state $p=p(\ro)$ of the unified model, instead of $\ro(a)$, we substitute the former into Eq. (\ref{31}) and obtain the energy density of each fluid
\be
\n{r12}
\ro_c=\frac{p+(1-\ga_x)\ro}{\Dga},  \qquad  \ro_x=\frac{-p+(\ga_c-1)\ro}{\Dga },
\ee
as a function of $\ro$. Hence, this unified cosmological model has been split into a two-fluid model with energy transfer. The interaction $Q$ is obtained after using the equation of state $p=p(\ro)$, the conservation equation $\ro'=-\ro-p$ and $\ro''=\ro+p-p'$ in the source equation (\ref{2}). Thus, the interaction  
\be
\n{qp}
Q=\frac{1}{\Dga}\left[\ga_c\ga_x\ro+(\ro+p)\left(1-(\ga_c+\ga_x)+\frac{dp}{d\ro}\right)\right],
\ee
becomes a function of $\ro$. 

We conclude that instead of considering interacting and unified models separately, it is better to see them as equivalent models. However, the conclusion could be different when the barotropic indices $\ga_c$ and $\ga_x$ are not constants. 

\subsection{The Chaplygin gas as an interacting two-fluid model}

For illustration, we consider the class of Chaplygin gases generated by the equation of state (\ref{pcha}) and express it as an interacting two-fluid model. To this end, we insert $p=p(\ro)$, given by the Eq. (\ref{pcha}), into Eqs. (\ref{r12})-(\ref{qp}) and find the dark matter and dark energy densities 
\ben
\n{r1cha}
\ro_{cCh}=\frac{\ro}{\Dga}\left[\ga_0-\ga_x-\frac{B}{\ro^{\al+1}}\right],\\
\n{r2cha}
\ro_{xCh}=\frac{\ro}{\Dga}\left[\ga_c-\ga_0+\frac{B}{\ro^{\al+1}}\right],
\een
and the interaction $Q$
$$
Q_{Ch}=\al\frac{\ga_c\ro_c\ro_x+\ga_x\ro_x^2}{\ro}
$$
\be\n{Qcha}
+(1+\al)(\ga_0-\ga_c)(\ga_c\ro_c+\ga_x\ro_x)+\ga_x\ro_x.
\ee
It can be written as a function of the total energy density $\ro$ by replacing the energy densities (\ref{r1cha})-(\ref{r2cha}) in (\ref{Qcha}) or as a function of the scale factor by using Eqs. (\ref{r}) and (\ref{r1cha})-(\ref{Qcha}). The interacting models, associated with the energy densities (\ref{r1cha})-(\ref{r2cha}) and the interaction (\ref{Qcha}), are parametrized by $\ga_c$ and $\ga_x$. It means that we have a set of  coupled two-fluid models which are related with a given  unified model and they are described by the effective equation of state (\ref{pcha}). However, these models differ by the ratio $r_{Ch}=\ro_{cCh}/\ro_{xCh}$
\be
\n{rcha}
r_{Ch}=\frac{-B+(\ga_0-\ga_x)\ro^{\al+1}}{B+(\ga_c-\ga_0)\ro^{\al+1}}.
\ee
In particular, for the simplest realization of the Chaplygin gas model we have investigated in the previous section (with $\ga_c=\ga_0$ and $\ga_x=0$), we obtain
\ben
\n{r12s}
&&\ro_{cCh}=\ro-\frac{B}{\ga_0\ro^{\al}},\quad \ro_{xCh}=\frac{B}{\ga_0\ro^{\al}},\\
\n{rs}
&&r_{Ch}=-1+\frac{\ga_0}{B}\,\ro^{\al+1},
\een
and the interaction (\ref{Qcha}) reduces to (\ref{ch}). On the other hand, by inserting the Eq. (\ref{rocha}) into the Eqs. (\ref{r1cha})-(\ref{rs}) we can express the relevant quantities characterizing the interacting two-fluid model in terms of the scale factor. 

For expanding universes and $\ga_0(1+\al)>0$, the total energy density (\ref{rocha}) $\ro^{\al+1}\to B/\ga_0$ and the ratio has the non-vanishing limit $r_{Ch}\to-\ga_x/\ga_c$ (see the Eq. (\ref{rcha})). Thus, the above decomposition of the Chaplygin gas could be used to alleviate the problem of coincidence. However, the simplest realization of the Chaplygin gas (\ref{rs}) has a vanishing limit for the ratio  $r_{Ch}\to 0$ and it is unusefulness to solve the problem of coincidence. Also, for $\al=\ga_0=1$, we get the original version of the Chaplygin gas.

\section{Conclusions}

We have investigated several models of dark matter and dark energy with energy transfer and shown that they can be considered as unified models, where both dark components are replaced by an effective one-fluid description with an effective equation of state. The two coupled equations describing the interacting model have been combined to obtain a second order differential equation for the total energy density, "the source equation". We have assumed a separable interaction $Q=\ro Q(\ga,\ga')$, which includes a large set of  cases investigated in the literature, and found the conditions of stability for the scaling solutions. 

We have presented the "linear interaction I" and imposed the conditions of stability to restrict the fourth constants of $Q_l=b_1\ro+b_2\ro_c+b_3\ro_x+b_4\ro'$. Then, we have found the two stationary solutions $\ga_s$ and $\ga_l^+$ of the evolution equation for the effective barotropic index, being $\ga_s$ the attractor solution. Interestingly, the existence of the attractor solution is linked to the fulfillment of the requirement $Q_l(\ga_s)<0$, indicating that the energy is being transfered from the dark energy to the dark matter. We have considered several particular examples by analyzing each term of $Q_l$ separately and found that some of these terms i.e., the total energy density, its derivative and the dark energy density are satisfactory couplings because in each case, the ratio $r_s=\ro_{cs}/\ro_{xs}$ is and attractor. These simple models may alleviate the problem of coincidence. Although, a coupling proportional to the dark matter energy density does not lead to stable solutions however, it can work when it is combined with the remaining terms of $Q_l$. 

Taking into account that $\ro_c$, $\ro_x$, $\ro_c'$, $\ro_x'$, $\ro$, $\ro'$ and $\ro''$ can be written as linear functions of the basis elements $\ro$, $\ro'$ and $\ro''$, we have investigated the "linear interaction II",  by introducing a linear combination of the former terms $Q_L=c_1\ro_c+c_2\ro_x+c_3\ro_c'+c_4\ro_x'+c_5\ro+c_6\ro'+c_7\ro''$, and reduced it to a linear combination of the latter ones $Q_L=c_1'\ro+c_2'\ro'+c_3'\ro''$. As far as we know, this interaction has not been investigated in the literature. We have imposed the conditions of stability and obtained the stationary solution for the effective barotropic index, $\ga_s$ and $\ga_L^+$, being $\ga_s$ the attractor solution, the exact scale factor and the effective barotropic index in implicit form. To generalize the above coupling, we have gone a step more by adding a constant and a well-behaved function of the scale factor to the "linear interaction II". Then, we have introduced the "general linear interaction", $Q_{gL}=\ga_s\ga_L^+\La_{eff}/\Dga+Q_L$, and a "more general linear interaction" $Q_{mgL}=\ga_L^+(\ga_s^2-v)\La_{eff}/\ga_s\Dga+f(\eta)/\Dga+Q_L$. Since the relative weight between the two first terms in $Q_{mgL}$ determines the final behavior of the scale factor, the models generated by those interactions have given several alternatives to the $\La$CDM model.

We have presented a class on "nonlinear interaction" $Q_{nL}=(c_1''\ro^{-1}\ro'^2+c_2''\ro+c_3''\ro'+c_4''\ro''+f(\eta)\ro^{c_1''/(1-c_4'')})/\Dga$, which leads to several important results. Although the source equation becomes a nonlinear differential equation, we have linearized and reduced it to the equation of motion for a forced linear oscillator with dissipative or antidissipative effects. Then, the analysis of this interaction has been separated into two main parts, the homogeneous case and the inhomogeneous one. In general, we have found that the equation of state of the effective one-fluid model depends explicitly on the scale factor. On the other hand, the universe evolves to a power-law scenario for large cosmological times. However for interactions having the form $Q_{nL}=\al\ga_c\ro_c\ro_x/\ro$ with $\al$ a constant, we have shown that the effective equation of state becomes that of the Chaplygin gas when the dark energy component is described by some kind of vacuum energy density i.e., $\ga_x=0$. Also, we have investigated the "reduced unified model" obtained for the particular nonlinear interaction leading to the source equation $x''=0$. In this case we have shown that the effective equation of state can be interpreted as a generalization of the polytropic equation of state.

Generically, when there are no restrictions on $Q_{nL}$, the equation of state of the effective one-fluid model defines what we have called the "relaxed Chaplygin gas model". It contains various generalizations of the Chaplygin gas, including the variable modified Chaplygin gas model with equation of state $p=A\ro+B(a)/\ro^\al$. To learn more about this nonlinear interaction, we have included a short analysis devoted to the structural stability of the source equation solutions. This has been made by establishing some connection between those solutions and the respective function $f(a)$ contained in the nonlinear interaction. For a selected set of functions $f(a)$, we have shown that the final asymptotic solutions of the source equation can be easily obtained.

Finally, we have given a prescription to obtain an interacting model starting form a unified one and found the energy densities of the dark components together with the interaction term that generates the former model. For illustration we have applied the prescription to the Chaplygin gas getting the relevant quantities which define the interacting model.

\acknowledgments

The author thanks the University of Buenos Aires for the partial support of this work during its different stages under Project X044, and the Consejo Nacional de Investigaciones Cient\'\i ficas y T\'ecnicas under Project PIP 114-200801-00328.

\end{document}